\documentclass[conference]{IEEEtran}
\IEEEoverridecommandlockouts
\usepackage{placeins} 
\usepackage{cite}
\usepackage{amsmath,amssymb,amsfonts}
\usepackage{algorithmic}
\usepackage{graphicx}
\usepackage{textcomp}
\usepackage{xcolor}
\usepackage{svg}
\usepackage{epsf,exscale,times}
\usepackage[english]{babel}
\usepackage{fancyhdr}
\usepackage{siunitx}
\usepackage{subcaption}
\usepackage{algorithm}
\usepackage[acronym]{glossaries}
\usepackage{hyperref}
\usepackage{ifthen}
\usepackage{textgreek}
\usepackage{ifthen}
\usepackage{gensymb}
\usepackage{tikz}
\usepackage{textcomp} 
\usepackage{mathtools}
\usepackage{microtype}
\usepackage{booktabs}  
\usepackage{multirow}
\newacronym{6dof}{6-DoF}{six degrees of freedom}
\newacronym{MI}{MI}{Mutual information}
\newacronym{MINE}{MINE}{Mutual Information Neural Estimation}
\newacronym{GP}{GP}{Gaussian process}
\newacronym{MLAT}{MLAT}{multilateration}
\newacronym{TOA}{TOA}{time of arrival}
\newacronym{UWB}{UWB}{ultra wideband}
\newacronym{PMF}{PMF}{probability mass function}
\newacronym{UE}{UE}{user element}
\newacronym{CIR}{CIR}{channel impulse response}
\newacronym{PSO}{PSO}{particle swarm optimization}
\newacronym{LOS}{LOS}{line of sight}
\newacronym{SGA}{SGA}{stochastic gradient ascent}
\newacronym{BN}{BN}{Batch Normalization}
\newacronym{ELU}{ELU}{Exponential Linear Unit}
\newacronym{CPU}{CPU}{central processing unit}
\newacronym{GPU}{GPU}{graphics processing unit}
\newacronym{RAM}{RAM}{random access memory}
\newacronym{EMA}{EMA}{exponential moving average}
\newacronym{DOP}{DOP}{dilution of precision}
\newacronym{SVD}{SVD}{singular value decomposition}
\newacronym{CRB}{CRB}{Cramér–Rao bound}
\newacronym{FIM}{FIM}{Fisher information matrix}
\newacronym{TWR}{TWR}{two-way ranging}
\newacronym{RMSE}{RMSE}{root-mean-square error}
\newacronym{AMR}{AMR}{autonomous mobile robot}
\newacronym{RNN}{RNN}{recurrent neural network}
\newacronym{LSE}{LSE}{log sum exponent}
\newacronym{RSS}{RSS}{received signal strength}
\newacronym{MVUE}{MVUE}{Minimum-variance unbiased estimator}
\newacronym{MAE}{MAE}{mean absolute error}
\newacronym{GAE}{GAE}{geometric average error}
\newacronym{dFoV}{dFoV}{diagonal field-of-view}
\newacronym{FoV}{FoV}{field-of-view}
\newacronym{RANSAC}{RANSAC}{random sample consensus}
\newacronym{GNSS}{GNSS}{global navigation satellite system}
\newacronym{IPS}{IPS}{indoor positioning system}
\newacronym{IR}{IR}{infrared}
\newacronym{LiDAR}{LiDAR}{light detection and ranging}
\newacronym{radar}{radar}{radio detection and ranging}
\newacronym{IMU}{IMU}{inertial measurement unit}
\newacronym{RF}{RF}{radio frequency}
\newacronym{PnP}{PnP}{Perspective-n-Point}
\newacronym{EKF}{EKF}{Extended Kalman filter}
\newacronym{SLAM}{SLAM}{simultaneous localization and mapping}
\newacronym{ODOP}{ODOP}{orientational dilution of precision}
\newacronym{DLT}{DLT}{direct linear transform}
\newacronym{MCP}{MCP}{multiple coverage probability}
\newacronym{EGA}{EGA}{elimination genetic algorithm}
\newacronym{DWNA}{DWNA}{discrete white noise acceleration}
\newacronym{LCD}{LCD}{liquid-crystal display}
\newacronym{SURF}{SURF}{speeded up robust features}
\newacronym{GG}{GG}{GroundGazer}
\newacronym{ToF}{ToF}{time of flight}
\newacronym{MIMO}{MIMO}{multiple input multiple output}
\newacronym{CSI}{CSI}{channel state information}
\newacronym{NLOS}{NLOS}{non-line-of-sight}
\newacronym{PVC}{PVC}{Polyvinylchlorid}
\newacronym{FMCW}{FMCW}{frequency-modulated continuous wave}
\newacronym{QR}{QR}{quick response}
\newacronym{FPGA}{FPGA}{field programmable gate array}
\newacronym{CDF}{CDF}{cumulative distribution function}
\newacronym{RCS}{RCS}{radar cross section}
\newacronym{SBR}{SBR}{shooting and bouncing rays}
\newacronym{AoA}{AoA}{angle of arrival}
\newacronym{AWGN}{AWGN}{additive white Gaussian noise}
\newacronym{LRP}{LRP}{Local Reference Point}

\usepackage{microtype}

\begin{document}
\normalsize
\title{GroundGazer: An optical reference system for planar localization with millimeter accuracy at low cost}
\author{\IEEEauthorblockN{Sven Hinderer, Jakob Hüsken, Bohan Sun, Bin Yang}
	\IEEEauthorblockA{\textit{Institute of Signal Processing and System Theory, University of Stuttgart, Stuttgart, Germany} \\
		firstname.lastname@iss.uni-stuttgart.de}
}

\maketitle

\noindent\textbf{Note:} This is the accepted version of the paper. 
The final version is published in the \emph{Proceedings of the IEEE 2026 15th International Conference on Indoor Positioning and Indoor Navigation (IPIN)}. 
DOI: \href{https://doi.org/10.1109/IPIN72673.2026.11772537}{10.1109/IPIN72673.2026.11772537}

\vspace{1em}

\renewcommand{\thefootnote}{}
\footnotetext{
	\textcopyright~2025 Personal use of this material is permitted.  Permission from IEEE must be obtained for all other uses, in any current or future media, including reprinting/republishing this material for advertising or promotional purposes, creating new collective works, for resale or redistribution to servers or lists, or reuse of any copyrighted component of this work in other works
}
\renewcommand{\thefootnote}{\arabic{footnote}}

\begin{abstract}
	
	Highly accurate indoor localization systems with absolute mm positioning accuracy are currently expensive. They include laser trackers, total stations, and motion capture systems relying on multiple high-end cameras. In this work, we introduce a high-accuracy, planar indoor localization system  named \gls{GG} for \glspl{AMR}. \gls{GG} estimates the AMR's planar position with mm and its heading with sub-degree accuracy. The system requires only a monocular (fisheye) camera, a chessboard floor, and an optional laser diode. Our system is simple and low-cost due to the chessboard floor, robust, scalable to multiple robots, and extendable to 3D position and orientation estimation.
	
\end{abstract}
\vspace{1ex}

\section{Introduction}
\label{sec:intro}
Benchmarking lower-accuracy indoor positioning systems with up to cm localization accuracy requires localization approaches with reliable mm accuracy. Currently, such systems (described in Sec.~\ref{sec:general_loc}) are very expensive. Costs for deploying a motion capture system, a common solution for accurate position referencing, can easily exceed $\$10000$, even for a small laboratory. While some alternatives might be cheaper, they usually come with other drawbacks such as strict \gls{LOS} requirements, or degradation with multi-path.

\subsection{Contributions} 
In this work, we introduce a camera-based, planar indoor localization system for \glspl{AMR} which achieves mm localization and sub-degree heading accuracy at multitudes lower cost than existing systems of similar precision for small rooms. We name it GroundGazer, as the camera placed on the \gls{AMR} gazes at the passive position references on the ground.
Its intended application is to serve as reference for less accurate localization methods. The proposed \gls{GG} prototype requires only a chessboard floor (or similar grid pattern), a camera (here a global shutter RGB fisheye camera), and an optional laser diode, i.e. low-cost, off-the-shelve components. The current implementation estimates planar position ($x$, $y$) and heading $\theta$, which is sufficient for planar \gls{AMR} localization with known or irrelevant height, roll, and pitch. In Sec.~\ref{sec:discussion}, we discuss an extension to 3D position and orientation estimation, at the cost of higher computational complexity and possibly reduced accuracy.

The idea for \gls{GG} originates from
\cite{ground_truth_laser}. To evaluate the accuracy of their proposed positioning system, they installed two point laser diodes that project two points from the front and the back of their localized vehicle onto the chessboard floor beneath it. The manually collected positions of those two points w.r.t. the known corner coordinates of the chessboard enable precise position and heading referencing. Our \gls{GG} system relies on the same idea, resulting in a highly accurate, automated,  camera-based indoor localization system. 

\subsection{System overview}
\begin{figure}[h]
	\centering
	\begin{subfigure}[t]{0.23\textwidth}
		\centering
		\includegraphics[width=\textwidth]{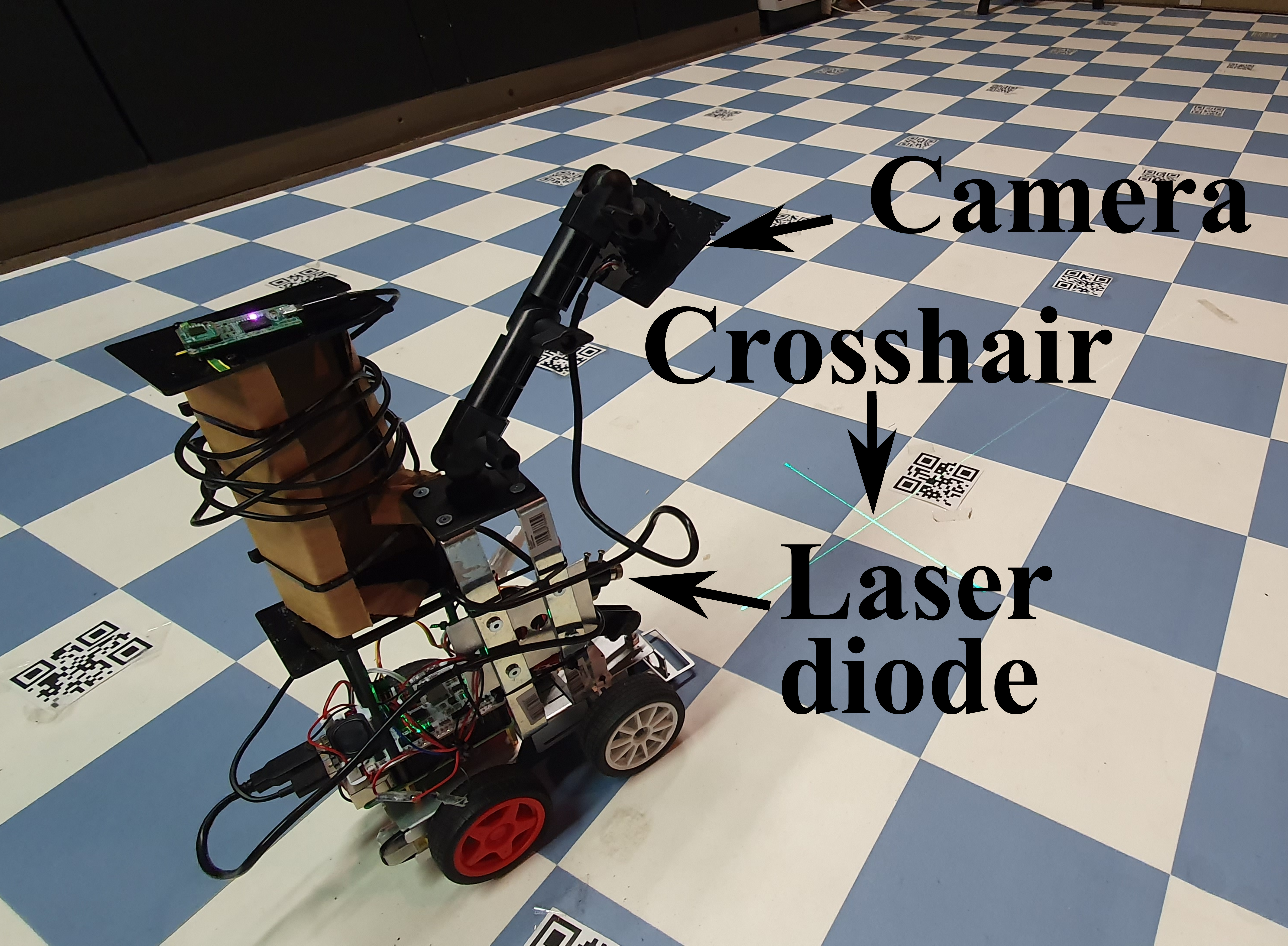}
		\caption{Our system in the laboratory equipped with a chessboard floor and \gls{QR} codes. The \gls{AMR} has a ground gazing camera and a laser diode that projects a crosshair on the floor at the front.}
		\label{fig:overview_system}
	\end{subfigure}
	\hfill
	\begin{subfigure}[t]{0.23\textwidth}
		\centering
		\includegraphics[width=\textwidth]{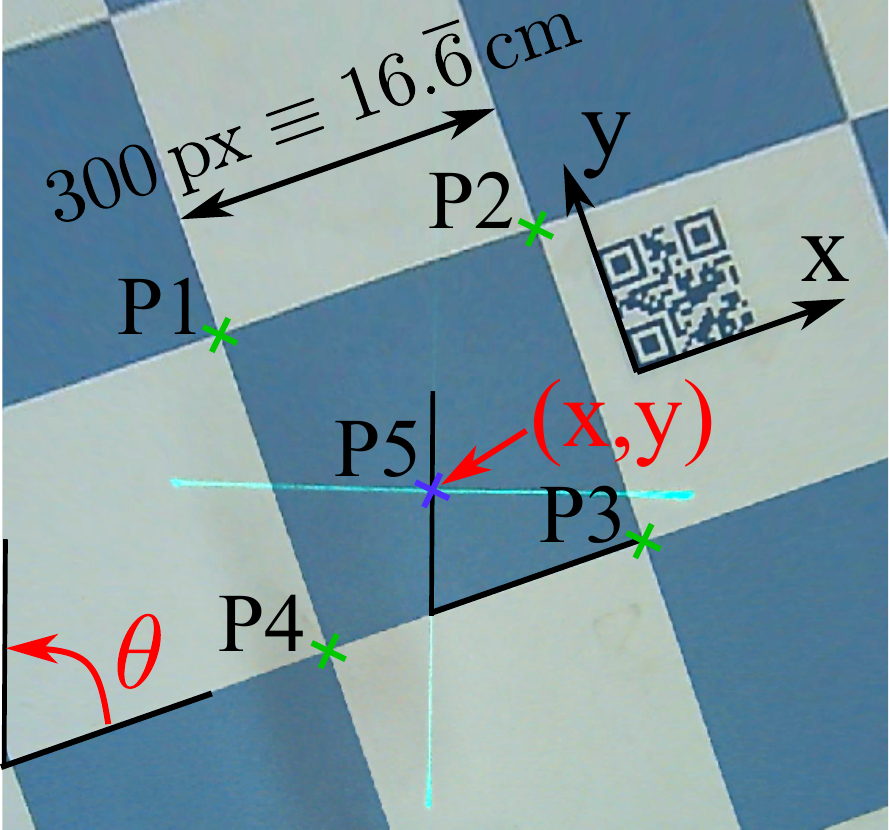}
		\caption{2D position and heading estimation using the detected crosshair (P5), the chessboard square corners (P1-P4), and the $x$-direction chessboard lines. The globally known square corner coordinates through the QR codes enable accurate, global positioning.}
		\label{fig:overview_localization}
	\end{subfigure}
	\caption{Overview of our GG localization system in a) and conceptual depiction of the localization method in b).}
	
	\label{fig:overview}
\end{figure}

The \gls{GG} prototype is depicted in Fig.~\ref{fig:overview_system}.
It consists of an \gls{AMR} with an RGB fisheye camera ($\sim \$100$) with large \gls{FoV} and global shutter. Global shutter cameras have lower motion blur than rolling shutter cameras due to parallel collection of all pixels. The camera is facing the ground and a green laser diode $(<\$10)$ projects a crosshair on said ground. The ground itself is made up of a chessboard pattern with well known grid size. Such floors can be purchased for around $\$$5-10 per $\mathrm{m}^2$. For large environments the chessboard floor can cause significant cost. While other systems like motion capture have higher cost for large rooms, it still limits the low-cost property of \gls{GG} to smaller rooms. However, for laboratories like ours, such chessboard floor might already be installed, reducing the localization system cost mainly to the cost of the camera. 

The localization system works by detecting the crosshair in a chessboard square. We selected a blue and white chessboard floor and a green laser diode, as green lasers are most easily visible and other commercially available black and white chessboard floors complicate the crosshair detection, caused by the large difference in laser reflective behavior between black and white squares. The camera image is calibrated such that the orientation of the chessboard in the image directly gives the heading. The corner coordinates of the chessboard squares are known and serve as passive position references for \gls{AMR} localization. 

Since the camera only observes a small, local area of the periodic chessboard pattern, the corner coordinates are ambiguous. Adding different \gls{QR} codes as global position and orientation references to the floor resolves this ambiguity. All \gls{QR} codes are coarsely aligned with the chessboard lines and their corners can thus be used to extract the $x$- and $y$-directions of the chessboard lines in global coordinates. The \gls{QR} code IDs allow for identification of the \gls{QR} code squares and global positioning. By measuring the offset between a detected \gls{QR} code square and the square containing the crosshair, the crosshair square and its known corner coordinates can be identified without requiring a \gls{QR} code in each square. 

The \gls{AMR} position is defined as the crosshair center location P5. Its planar position $(x, y)$ is given by the relative position of the crosshair w.r.t. its square corners P1-P4 and the heading $\theta$ is given by the angle between the chessboard lines in $x$-direction and the vertical image axis. An image describing the \gls{AMR} positioning is shown in Fig.~\ref{fig:overview_localization}.

\gls{GG}'s system design enables high accuracy with simple visual infrastructure. While the requirement of a chessboard floor limits \gls{GG}'s use as a general localization method, deploying such chessboard floor comes at reasonable effort and at low cost for small areas. \gls{GG} is not restricted to visible light and could also be realized with an active \gls{IR} camera and a passive \gls{IR} grid, improving robustness to ambient illumination and allowing the positioning infrastructure to remain largely unobtrusive for human users. Our main contribution therefore lies in the design of the localization system rather than in the image processing used to estimate the \gls{AMR}'s position and heading, as described in Sec.~\ref{sec:ground_gazer}. The latter primarily relies on openly available OpenCV functionality \cite{opencv_library}.

\section{Related works}
We focus on closely related systems in this section. For an overview of general indoor localization technologies, we refer to the surveys in \cite{review_indoor_pos_habil} and \cite{review_indoor_loc}, and for optical indoor localization systems to \cite{Survey_Optical_pos} and \cite{cv_loc_survey}.
\subsection{Related works with ground gazing cameras}
Infrastructure-free approaches localize using only the given ground texture. Relative positioning can be realized through \gls{SLAM} or feature-matching in successive images to build a vision-based odometer \cite{ground_visual_odometer}. Absolute positioning is possible with feature-matching of images collected during inference with a pre-collected image database of the ground \cite{high_precision_matching, feature_matching_newest_GNN, feature_matching_icra25}. However, mapping a floor and anchoring it to the real world with correct absolute scale requires an expensive measurement campaign with accurate ground truth, e.g. from motion capture systems \cite{feature_matching_newest_GNN}.

While feature-matching requires no infrastructure, using known position references is the more suitable approach for reference systems due to higher accuracy, robustness, easier setup, and lower hardware requirements. Various systems place infrastructure (fiducial markers) on the floor for absolute, global localization of \glspl{AMR} \cite{ground_april} and other robots, such as walking assistants \cite{QR_assistant} and drones \cite{QR_drones}. By applying popular markers including QR codes, ArUco markers, or AprilTags, they achieve dm to cm localization accuracy. A QR code-based docking system in \cite{QR_docking_submm} reports sub-mm accuracy with very small camera-QR code distance. Although low-cost, the very small camera-\gls{QR} code distance and the need for many \gls{QR} codes limit its use for \gls{AMR} localization. Visual floor marker grids for \gls{AMR} navigation can also be found in commercial systems, e.g. the Hercules \gls{AMR} by Amazon robotics. 

The chessboard line grid in \gls{GG} is easier to realize than a precisely placed marker grid and offers more accurate corner detection with high robustness against occlusion due to its redundant pattern. In a laboratory, it can also facilitate installation of other equipment.
\subsection{Localization technologies with mm positioning accuracy}
\label{sec:general_loc}
Optical systems include motion capture systems like the commercial systems by Qualisys and Vicon\footnote{Available at: \href{https://www.qualisys.com/}{https://www.qualisys.com/} and \href{https://www.vicon.com/}{https://www.vicon.com/}.}. They consist of passive \gls{IR} markers placed on the tracked \gls{AMR}. With one marker, 3D positioning is possible. A minimum of three non-collinear markers enable 3D position and orientation estimation. While capable of sub-mm 3D localization and sub-degree orientation estimation, they require \gls{LOS}, multiple high-end \gls{IR} cameras, and become very expensive for large-scale deployments. Further, since the cameras are not installed around the \gls{AMR}, accurate time synchronization is required to ensure mm accuracy. A different optical solution are fiducial markers. In \cite{ODOP}, sub-mm accuracy is achieved, but only with costly cameras and metrology-grade equipment to precisely install custom reference markers. The previously mentioned QR code docking system in \cite{QR_docking_submm} also gives sub-mm accuracy, but is limited by camera-QR code distance.

Total stations (tachymeters) pose another solution for 3D position estimation with mm or sub-mm accuracy, e.g. as used in \cite{esparagos}. Total stations enable long range sensing, but like motion capture systems, they are expensive and require \gls{LOS} and accurate synchronization.

In acoustic ultrasound systems, the low propagation speed of sound greatly reduces sensitivity to timing and synchronization errors compared to \gls{RF}-based \gls{UWB} systems. Using \gls{ToF}-based measurements between multiple active anchors and the tracked user, a commercial system reports real-time 3D positioning accuracy of up to 1.5 mm using 6 acoustic anchors with ideal geometric placement \cite{zerokey}.

While \gls{ToF}-based localization with \gls{UWB} systems only gives dm to cm accuracy, recent studies show mm accuracy by combining initial, coarse \gls{ToF}-based range estimates with refined, ambiguity-resolved phased-based range estimates \cite{UWB_mm}. Other \gls{RF}-based systems include the interferometry-based system developed by Koherent\footnote{Available at: \href{https://koherent.io/}{https://koherent.io/}}, and \gls{FMCW} radar-based backscatter systems using reflective tags that modulate the reflected signal \cite{hawkeye, mm_siemens}. Deep learning-based \gls{CSI} fingerprinting with a massive \gls{MIMO} system could recently also achieve mm accuracy in a small area and under \gls{LOS} conditions \cite{csi}. However, domain shifts between training and inference, caused by environmental changes, reduce the accuracy of such systems. \gls{RF}-based systems often require specialized hardware and degrade under multi-path. At smaller wavelength, using \gls{LiDAR} with reflectors at known positions as references has also shown mm accuracy \cite{lidar_mm_reflector}, but \gls{LiDAR} has high sensor cost.  

Additional systems with (sub-)mm accuracy based on visual markers, laser projections, magnetic fields, laser trackers, pseudolites, and infrared laser beams (iGPS)  can be found in \cite{Survey_Optical_pos} and \cite{mautz_alternative_systems}.

\section{GroundGazer}
\label{sec:ground_gazer}

\subsection{Image transformation}
\begin{figure}[h!]
	\centering
	\begin{subfigure}[b]{0.15\textwidth}
		\centering
		\includegraphics[width=\textwidth]{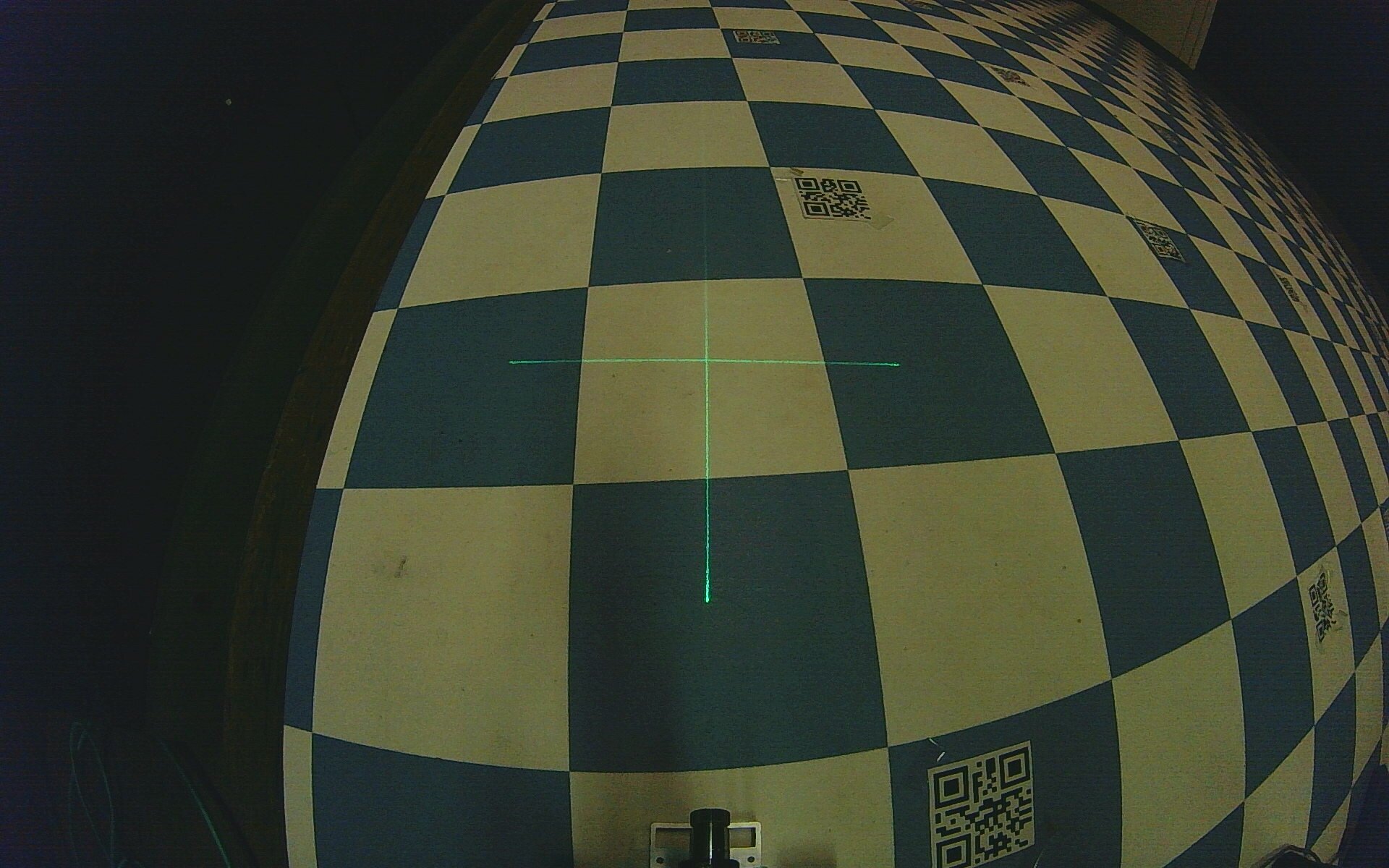}
		\caption{Original image.}
		\label{fig:original}
	\end{subfigure}
	\hfill
	\begin{subfigure}[b]{0.15\textwidth}
		\centering
		\includegraphics[width=\textwidth]{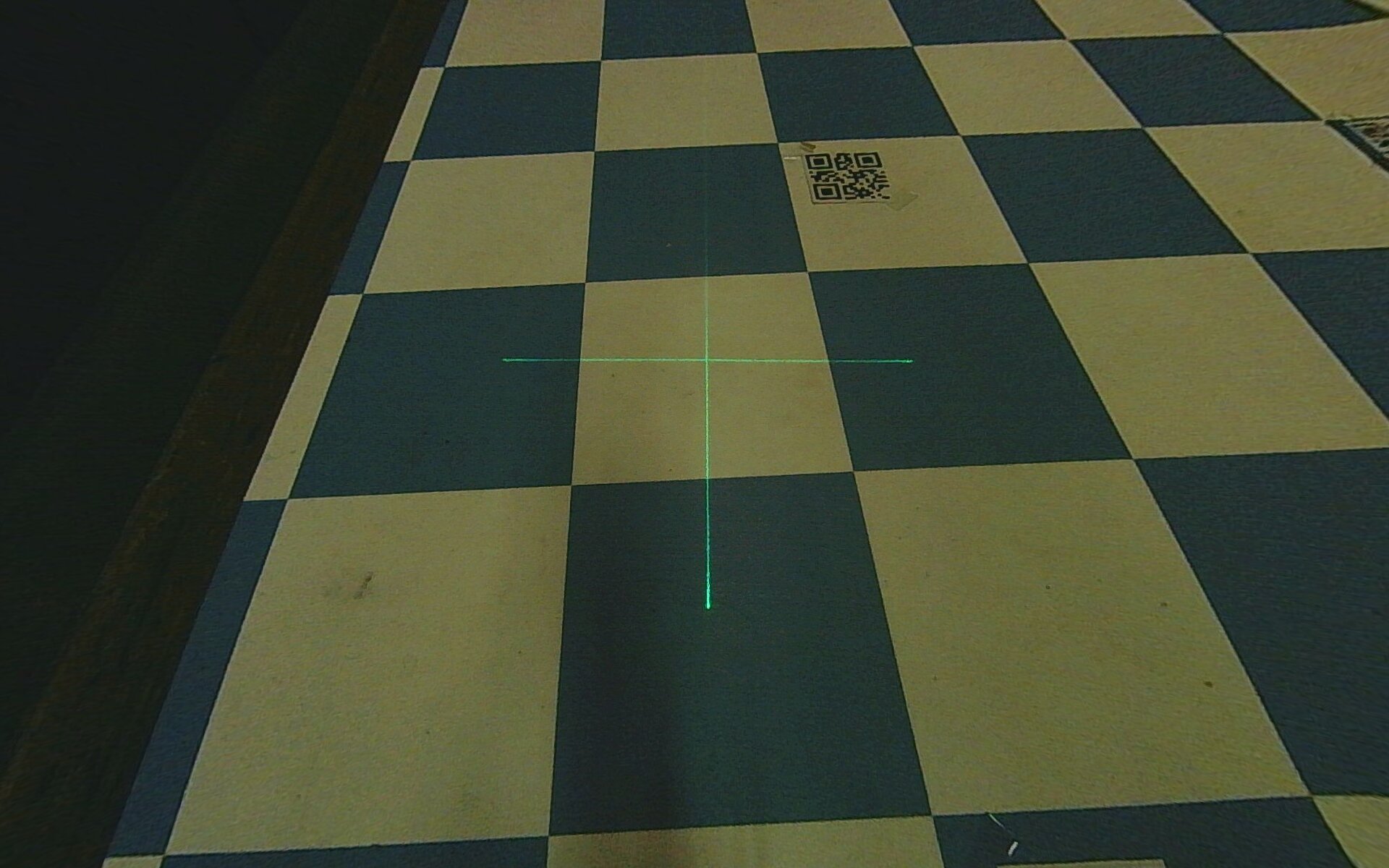}
		\caption{Undistorted.}
		\label{fig:undistorted}
	\end{subfigure}
	\hfill
	\begin{subfigure}[b]{0.15\textwidth}
		\centering
		\includegraphics[width=\textwidth]{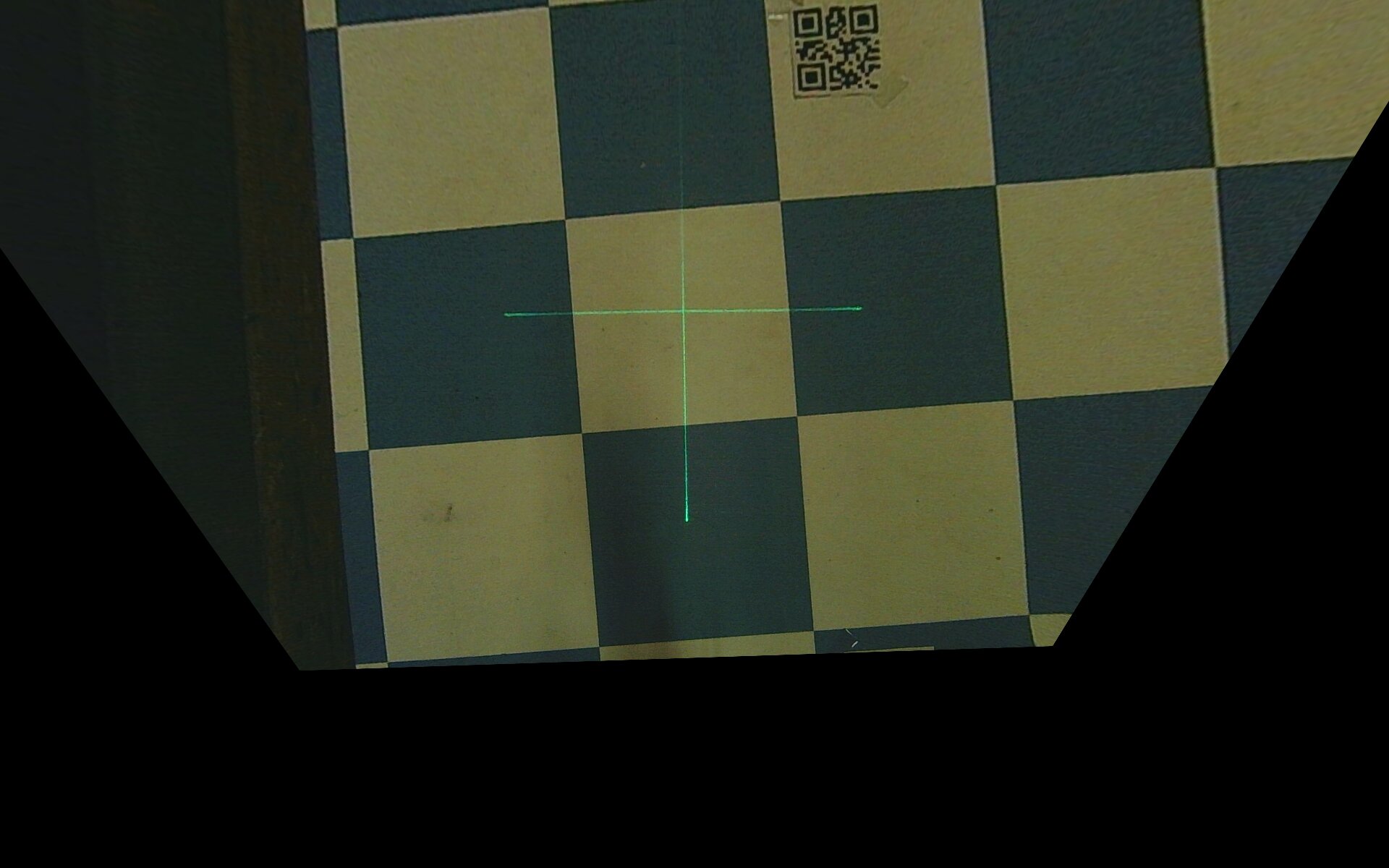}
		\caption{Top-down view.}
		\label{fig:topdown}
	\end{subfigure}	
	\caption{Transformation of an original image in a) through undistortion in b) and perspective transformation in c).}
	\label{fig:top_down_transform}
\end{figure}

The initial step of our \gls{GG} localization algorithm transforms the distorted and skewed image collected by the fisheye camera to a top-down view of the chessboard floor. This step consists of two transformations. First, the fisheye image is undistorted. The required intrinsic matrix $\mathbf{K}$ of the camera and its distortion coefficient vector $\underline{d}$ are estimated with a printed chessboard pattern. A distorted fisheye image and the undistorted result are shown in Fig.~\ref{fig:original} and Fig.~\ref{fig:undistorted}, respectively. Then, the undistorted image is transformed to a top-down view by perspective transformation, depicted in Fig.~\ref{fig:topdown}. Since \gls{GG} localizes the crosshair center relative to its square corners, errors in undistortion and perspective transformation have little influence on the positioning accuracy. The homography matrix for perspective transformation is estimated offline with four corresponding square corner points of both views. The \gls{AMR} heading has to match the orientation of the transformed chessboard to align them. We calibrate at $\theta=90^\circ$, thus the edges of the top-down view square are parallel to the image axes. We set its size to (300, 300) pixels. The four corresponding points from the undistorted image view are the corners of the central crosshair square.

Alternatively, heading could be estimated with the angle between a similarly calibrated, vertical crosshair line that points in heading direction and the $x$-direction chessboard line. This makes heading estimation independent of the camera orientation and thus of camera vibrations. However, it is still susceptible to lesser laser diode vibrations and it makes heading estimation dependent on the crosshair line detection, which is worse than the chessboard line detection due to broader edges, crosshair distortion on uneven ground etc., and therefore not chosen.

\subsection{\gls{QR} code detection and masking}
\gls{QR} code detection gives the four corner points of a \gls{QR} code in known order and its ID, which enables identification of all chessboard squares and their corners.

As the \gls{QR} codes would disturb successive image processing, we mask them out. Our procedure might also mask parts of the crosshair if it lies on a \gls{QR} code and could be improved in future work. Nevertheless, our experiments have shown that \gls{GG} can cope with irregular missed crosshair detections and by proper parametrization of the Hough transform in the crosshair detection, we can also detect an incomplete crosshair. 
\subsection{Crosshair detection}
\begin{figure}[h]
	\centering
	\begin{subfigure}[t]{0.23\textwidth}
		\centering
		\includegraphics[width=\textwidth]{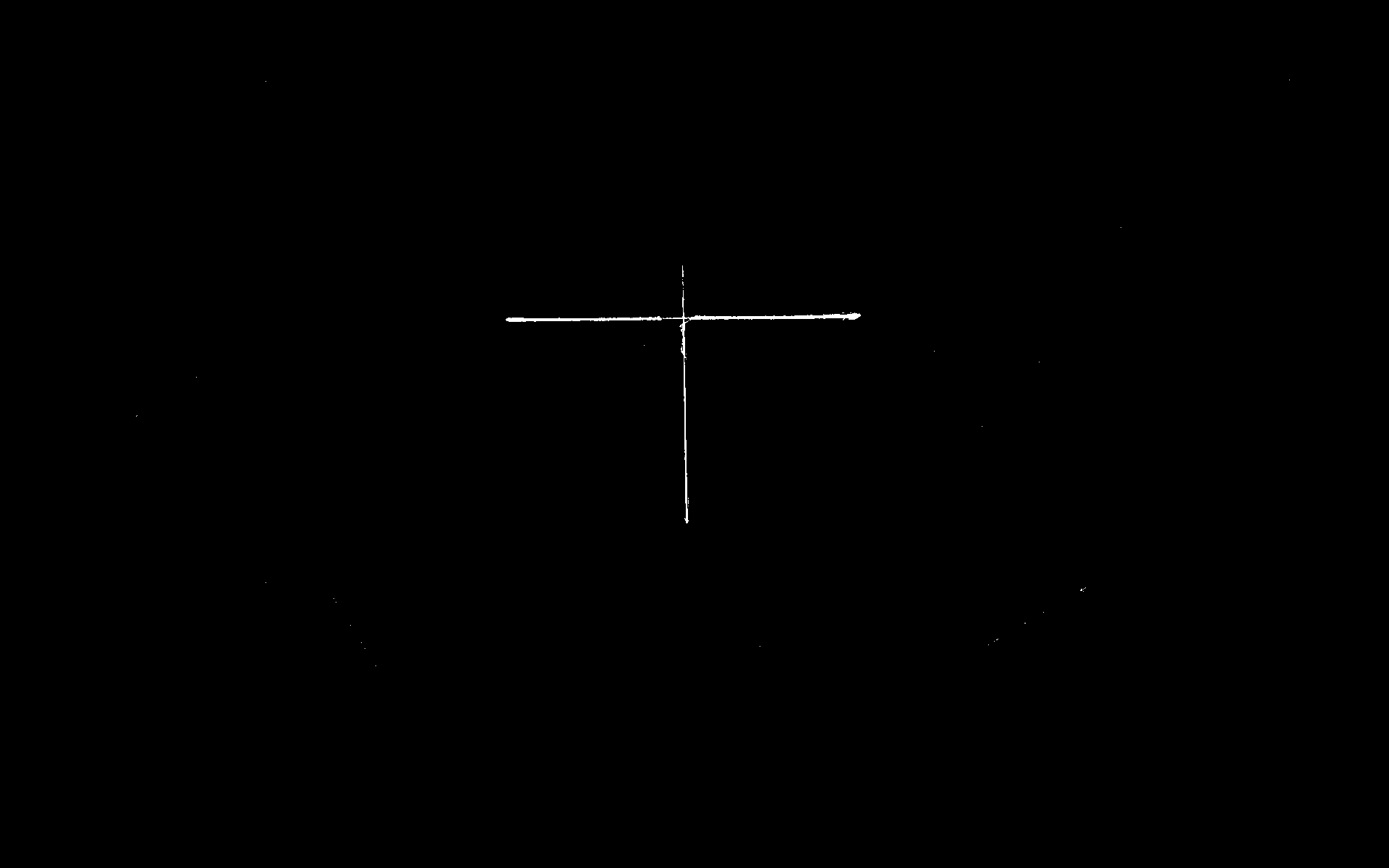}
		\caption{Crosshair mask.}
		\label{fig:crosshair_mask}
	\end{subfigure}
	\hfill
	\begin{subfigure}[t]{0.23\textwidth}
		\centering
		\includegraphics[width=\textwidth]{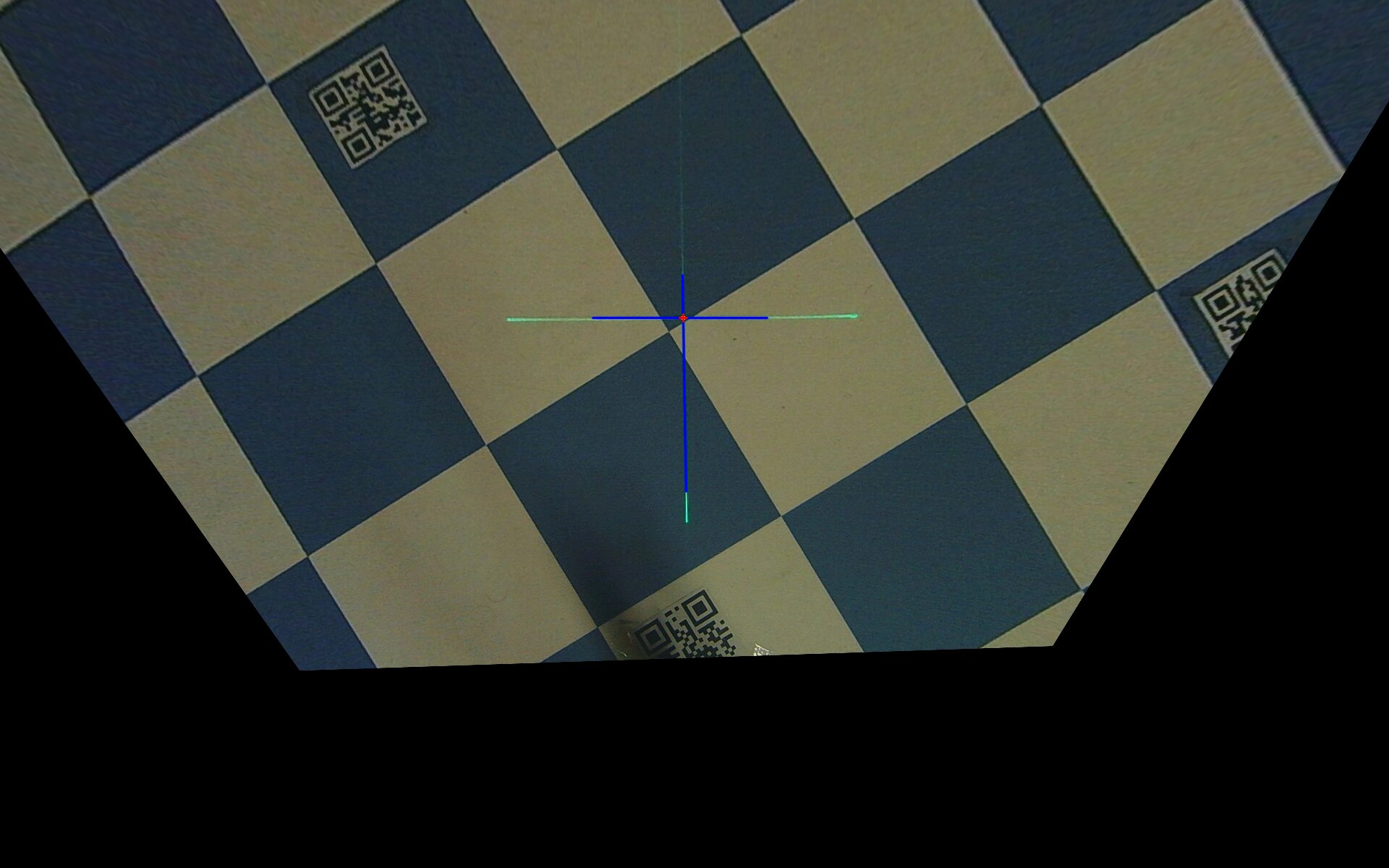}
		\caption{Final crosshair detection.}
		\label{fig:crosshair_lines_filtered}
	\end{subfigure}	
	\caption{Crosshair detection.}
	\label{fig:crosshair_detection}
\end{figure}

The green crosshair can be extracted by filtering the image with a mask that keeps the green image parts. This mask has to be carefully tuned, as both the different colors of the chessboard and varying illumination condition influence the detected crosshair color. We therefore apply a concatenation of multiple masks. The first mask filters out all pixels with low green values. We then convert the image to the HSV color space, which was more robust than only working in the RBG space. Three additional masks filter out pixels with low saturation, low hue, and pixels where the sum of hue and value is low. A combined crosshair mask is found in Fig.~\ref{fig:crosshair_mask}. The horizontal and vertical lines in a region of 150 pixels around the estimated crosshair position from the previous time step are then found by probabilistic Hough transformation, clustered into lines with similar angle and start- and endpoints, and the start and end $x$-values of the clustered lines are averaged for the vertical, and the $y$-values for the horizontal line. The crosshair location in image coordinates is then estimated as the intersection of the horizontal and vertical crosshair lines. The crosshair lines and the detected crosshair position are shown in Fig.~\ref{fig:crosshair_lines_filtered}.

If the crosshair is not detected or is detected at an unexpected position, we use an average of the previous crosshair positions as estimate. This case does not occur in the experiments. In principle, our prototype can work without the laser diode by replacing the real crosshair with a virtual one, which stays in the same position in the image under ideal conditions (i.e. with a perfectly stabilized camera). This means selecting one point in image coordinates to replace the crosshair center and using it as \gls{AMR} position. Heading is estimated based on the chessboard rather than the crosshair orientation. On our low-cost robot, the lower mounted laser diode proved to be more accurate and robust than a virtual crosshair, being less affected by vibrations under \gls{AMR} motion.

\subsection{Chessboard square detection}
\begin{figure}[h]
	\centering
	\begin{subfigure}[t]{0.155\textwidth}
		\centering
		\includegraphics[width=\textwidth]{Top_down}
		\caption{Top-down view.}
		\label{fig:original2}
	\end{subfigure}
	\hfill
	\begin{subfigure}[t]{0.155\textwidth}
		\centering
		\includegraphics[width=\textwidth]{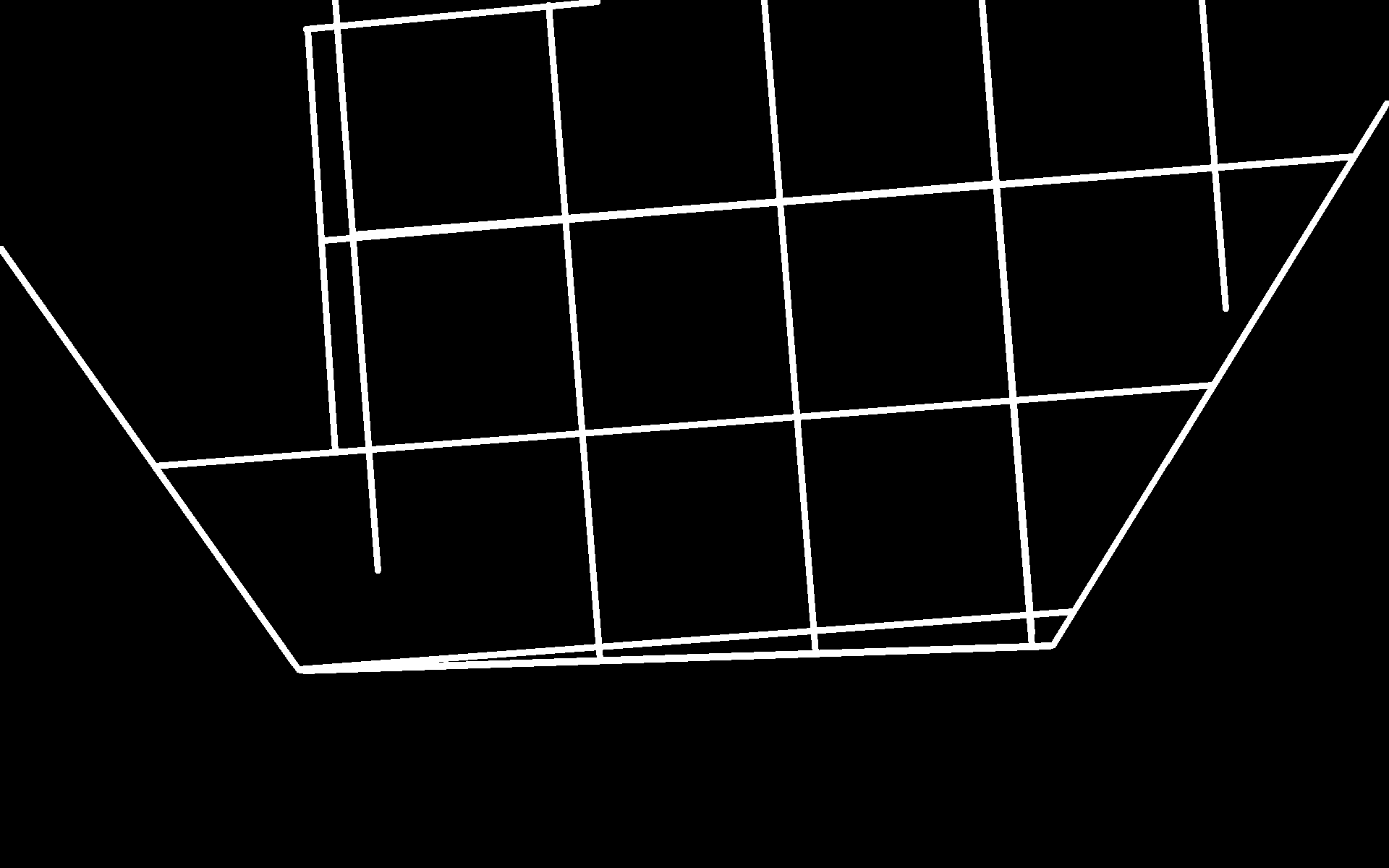}
		\caption{Hough lines after opening.}
		\label{fig:hough}
	\end{subfigure}	
	\hfill
	\begin{subfigure}[t]{0.155\textwidth}
		\centering
		\includegraphics[width=\textwidth]{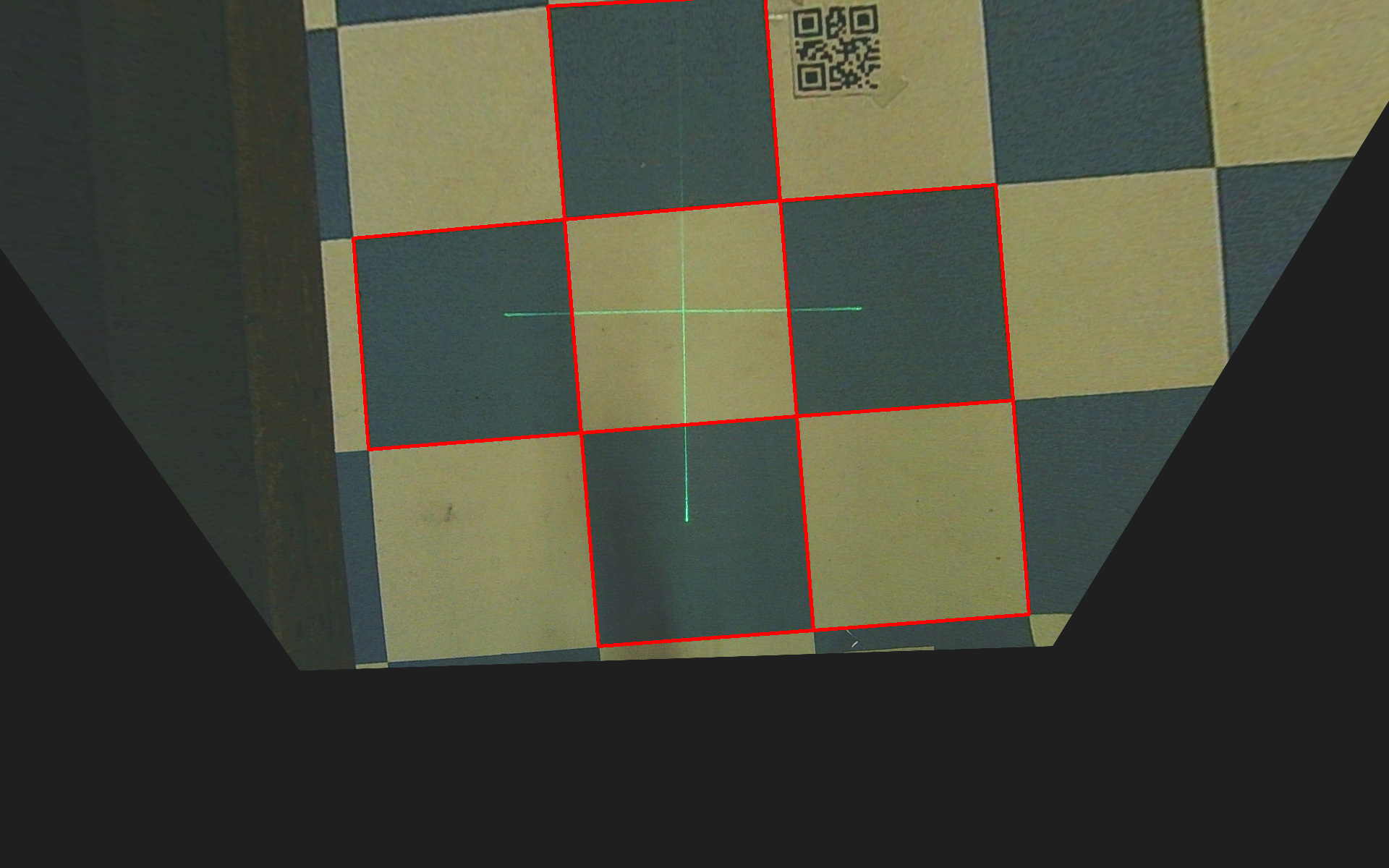}
		\caption{Square detection (red).}
		\label{fig:squares_refined}
	\end{subfigure}	
	\caption{Square detection in the top-down image.}
	\label{fig:top_down_squares}
\end{figure}

To detect the chessboard squares, we first apply a Canny filter for edge detection after grayscale conversion and Gaussian smoothing. The binary Canny output is then processed by probabilistic Hough transformation to find the chessboard lines. For further denoising, the detected lines are broadened and the image of lines is processed by an opening operation. The image after Hough transformation and opening is depicted in Fig.~\ref{fig:hough}. We then invert the image. The chessboard squares are detected by first finding and simplifying the white square contours. The simplified contours (polygons) are then filtered for square detection, where only polygons with four edges, relative side lengths between factor 0.95 and 1.05, and an area of 80.000 - 100.000 pixels are kept. Finally, the detected corners are refined to sub-pixel accuracy \cite{subpix}, shown in Fig.~\ref{fig:squares_refined}. 

If squares are occluded, the redundancy in the chessboard pattern enables estimating all squares by simply extending the detected squares to build virtual, undetected squares. Using the precise square detection, this method has proven quite accurate. If for some reason no square is detected, we apply a constant velocity model for $x$, $y$ and $\theta$, which is sufficient with a high enough sampling frequency of camera images. A full failure never happens during the experiments, but using a virtual crosshair square is required a few times when the lines of the crosshair merge with the lines or corners of the crosshair square, leading to chessboard corner detection failures.

\subsection{Position and heading estimation}
Given the detected corner points of the chessboard square with detected laser crosshair, we compute the relative position of the crosshair in the square. We then convert the crosshair pixel coordinates to global coordinates with the known square size of $16.\overline{6}$ cm. To resolve the rotation ambiguity of the local square coordinate system and compute global world coordinates, the \gls{QR} code references are used, as previously shown in Fig.~\ref{fig:overview_localization}. The heading is computed as the angle between the chessboard lines in $x$-direction and the calibrated vertical image direction. 

If no \gls{QR} code is detected, we predict the current heading using a constant velocity model. The heading is then estimated by choosing the detected angle from the set of four ambiguous angles (due to the lack of knowledge which chessboard lines point in $x$- and $y$-direction) by a nearest neighbor assignment. The four ambiguous chessboard line directions are computed by clustering and averaging the lines from Hough transformation in Fig.~\ref{fig:hough}. Since the \gls{AMR} has to move in a physically plausible way and the heading cannot make large jumps between two images, this nearest neighbor assignment works reliably.

\gls{AMR} position estimation is also possible without \gls{QR} codes through crosshair tracking in a simple fashion. Since we can reliably estimate the heading, no position ambiguity in local square coordinates occurs while tracking the crosshair, as the order of P1-P4 in Fig.~\ref{fig:overview_localization}, i.e. the orientation of a locally observed square, stays known. To catch jumps between squares, we check if the crosshair position in local square image coordinates in $x$ or $y$ surpasses a threshold of $150$ pixels and if so, we perform a global square jump in the respective direction.

Using this crosshair tracking after an initial global position and heading estimate with a QR code, further frequent global updates with the QR code references are not necessary, but this requires a high enough image sampling frequency such that the Nyquist-Shannon sampling theorem holds. The spatial frequency $f_{pattern}$ of the chessboard pattern is $f_{pattern} = \frac{1}{16.\overline{6}}\, \mathrm{cm}^{-1}$ and thus the images have to be sampled with a spatial frequency of at least $f_{image}=2f_{pattern}$ to avoid aliasing (causing jumps into the wrong square during crosshair tracking). This upper bounds the maximum \gls{AMR} velocity to $v_{max} = f_{image}^{-1}f_{fps}$, where $f_{fps}$ is the sampling rate of the camera. We record with $f_{fps} = 8.\overline{3}$ Hz, i.e. $v_{max} \approx 0.69\,\frac{\mathrm{m}}{\mathrm{s}}$, which is more than double of the maximum velocity of our \gls{AMR}. To adapt the crosshair tracking for faster \glspl{AMR}, one would have to either increase the unambiguous region (e.g. by increasing the square size), increase the sampling rate of the camera, or integrate an accurate kinematic model (e.g. through odometry).

If no QR code is detected in the first frames, we assume an initial position and heading of $[x,y,\theta]^T=[0,0,0]^T$ and then correct the initial estimates through rotation and translation once a QR code becomes available.

%

\section{Experiments}
\subsection{Hardware}
Our prototype \gls{AMR} is a low-cost Raspberry Pi robot (SunFounder Picar-X $\sim\$100$, Raspberry Pi 4b $\sim\$40$) with Ackermann kinematics. The USB RGB fisheye camera is from Svpro $(\sim\$100)$ and offers a resolution of 1920x1200 pixels, a \gls{FoV} of $>150^\circ$, and fast image creation (i.e. negligible motion blur) by employing a global shutter. The green laser diode is from CTRICALVER $(<\$10)$ with a wavelength of \SI{520}{nm} and draws $<$\SI{1}{mW} of power. The chessboard floor is made from \gls{PVC} with a known square size of ($16.\overline{6}$, $16.\overline{6}$) cm. The laboratory is illuminated by artificial lights located on one side of the ceiling, resulting in a well-lit but non-uniformly illuminated floor.

\subsection{Localization experiments}
To showcase the correct working of our system under real inference conditions with possible motion blur, \gls{AMR} vibrations etc., we drive the \gls{AMR} in the laboratory by steering it wirelessly with a PS4 controller while collecting the images. The AMR speed is around $\SI{20}{\frac{cm}{s}}$. The slow AMR speed in the experiment comes from the prototype AMR whose motion becomes notably non-smooth for higher speeds and is not an inherent GG limitation. Since we have no access to a costly high-accuracy reference system, which was our motivation to design the low-cost and high-accuracy \gls{GG}, we developed a software tool to label the ground truth position and heading based on the corresponding top-down image by hand, similar to the annotation procedure of various localization approaches based on pixel differences \cite{high_precision_matching, feature_matching_icra25}. We therefore zoom into the wanted points and select the crosshair center position, and three corners of the respective square to build the coordinate system for localization, including the $x$-direction line for heading estimation. The points are selected to the best of our abilities with pixel accuracy, where one pixel equals roughly $\frac{16.66}{300}\approx0.56$ mm (see Fig.~\ref{fig:overview_localization}). Future work should benchmark \gls{GG} against external ground truth, potentially in a version more robust to \gls{AMR} vibrations discussed in Sec.~\ref{sec:discussion}. All images are processed offline, as our goal is to use the prototype for offline position referencing.

\section{Results}
\begin{figure}[h]
	\centering
	
	\includegraphics[width=.48\textwidth]{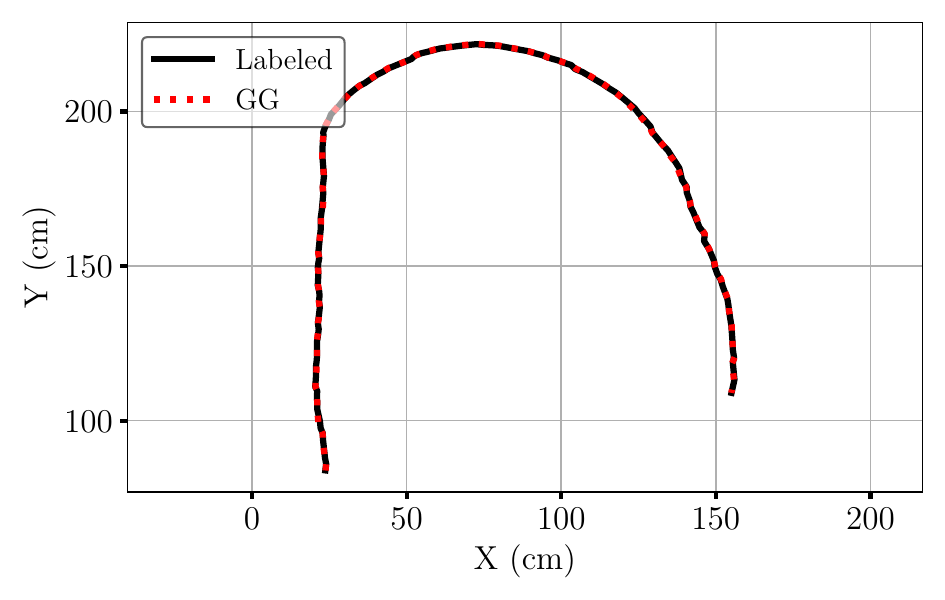}
	\caption{Trajectory evaluation.}
	\label{fig:trajectory}
	
	
	\label{fig:result}
\end{figure}

The labeled and estimated trajectories are shown in Fig.~\ref{fig:trajectory}. The estimates closely follow our hand-labeled trajectory. 

\begin{figure}[h]
	\centering
	\begin{subfigure}[t]{0.46\textwidth}
		\centering
		\includegraphics[width=\textwidth]{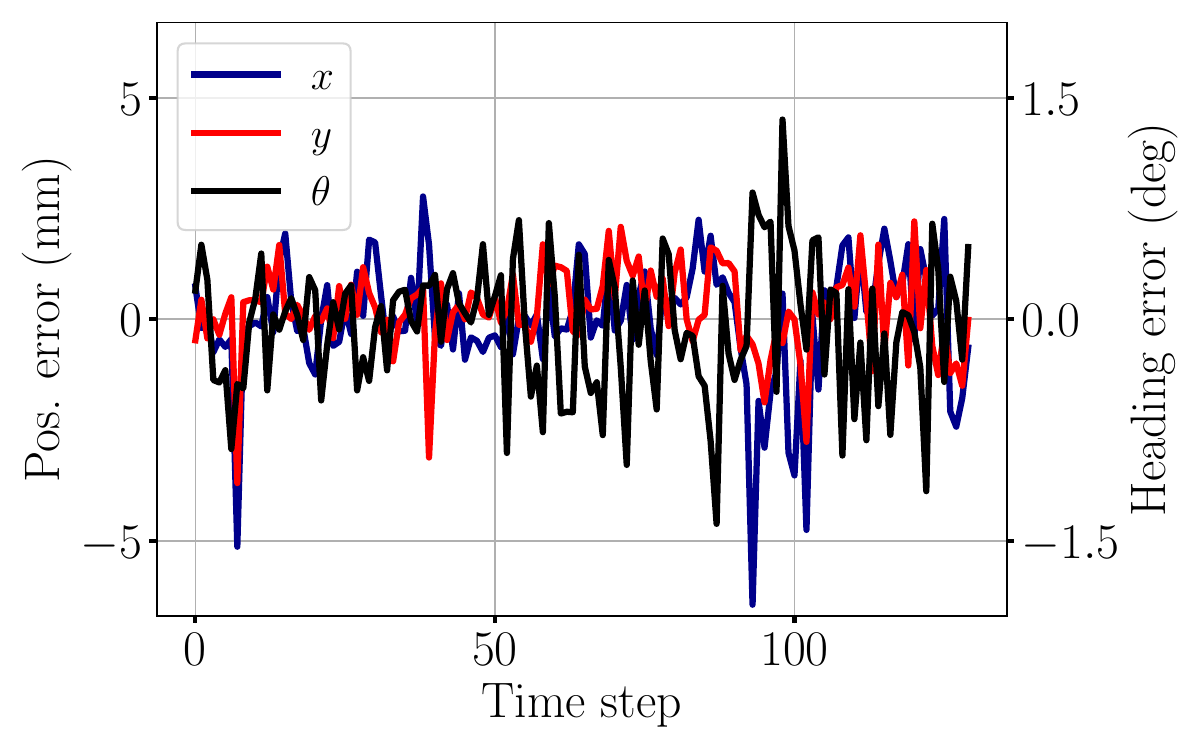}
		\caption{Errors.}
		\label{fig:errors}
	\end{subfigure}	
	\hfill
	\begin{subfigure}[t]{0.46\textwidth}
		\centering
		\includegraphics[width=\textwidth]{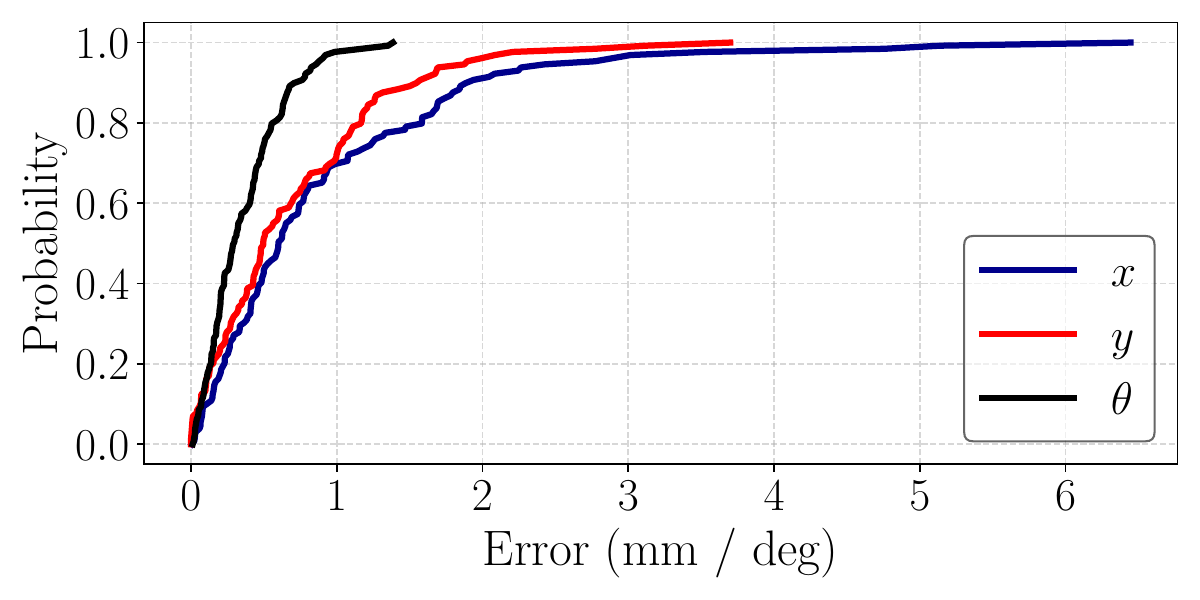}
		\caption{Cumulative distribution functions of the absolute errors.}
		\label{fig:hist}
	\end{subfigure}	
	
	\caption{Position and heading errors in a) and corresponding \glspl{CDF} of absolute errors in b).}
	\label{fig:results}
\end{figure}
A closer error evaluation is found in Fig.~\ref{fig:results}. The errors in $x$, $y$ and $\theta$ are depicted in Fig.~\ref{fig:errors}. A few outliers arise due to small errors in the crosshair or corner detection, often caused by overlaps of the crosshair with the lines and corners of the chessboard.  The corresponding CDFs of the absolute errors are found in Fig.~\ref{fig:hist}. They show clearly that most errors in $x$ and $y$ are in low mm or even sub-mm range.

Table.~\ref{tab:errors} gives the bias, the \gls{MAE}, the maximum absolute error (Max), and the 95th and 99th percentiles of the absolute error for each parameter.
\begin{table}[h]
	\centering
	\caption{Error statistics for $x$ and $y$ in mm and for $\theta$ in degrees.}
	\label{tab:errors}
	\begin{tabular}{l | r r r r r}
		\toprule
		& Bias & MAE & 95th & 99th & Max\\
		\midrule
		$x$         & -0.03 & 0.92 & 2.62 & 5.03 & 6.44\\
		$y$         & 0.02 & 0.69 & 1.89 & 3.02 & 3.70 \\
		$\theta$    & -0.05 & 0.36 & 0.87 & 1.30 & 1.39\\
		\bottomrule
	\end{tabular}
\end{table}
All estimates have negligible bias. Positioning errors in both dimensions are similar as expected. The errors in $x$ are higher due to more outliers as visible in Fig.~\ref{fig:errors}. The 99th percentile of the absolute errors and the max errors in $x$ and $y$ are well below 1 cm. Heading estimation also shows good performance with the 95th percentile of absolute heading errors being  slightly under $1^\circ$. Good heading accuracy is important to accurately map from the crosshair center to another position around the crosshair, e.g. to a localization sensor on the \gls{AMR}.

\section{Discussion}
\label{sec:discussion}
Our proposed \gls{GG} system is low-cost and gives mm accuracy. It is, however, just a prototype and this localization approach brings several advantages and can be further improved upon.

\textbf{Cost and portability:}
The largest cost factor of \gls{GG} is the chessboard floor, which is still cheaper than using other mm-accuracy methods for small rooms. Rooms without a chessboard floor could instead use a taped or drawn grid of lines. By connecting pre-built chessboard patches, one could also port the system.

\textbf{Accuracy:} Positioning accuracy is mostly independent of the \gls{AMR} position in the room, given similar illumination conditions. Current localization requires four crosshair square corners and might be extended to accept any (and less) corners. This allows crosshair projection closer to the \gls{AMR}, or replacing the crosshair center with a known, visible point on the \gls{AMR} (see e.g. Fig.~\ref{fig:original} bottom) for position referencing. The error in mapping from GG's position to a sensor on the \gls{AMR} could also be minimized with a camera mount beneath the \gls{AMR}. However, this reduces \gls{FoV} and the feasible chessboard square size. It also increases motion blur, necessitating reduced exposure time and thus additional lighting or camera flash. 

The current limitation during motion is the mechanical stability of the \gls{AMR}, causing camera and crosshair jitter. Since we only evaluate the crosshair position accuracy, not a jittered projection to a stable position on the \gls{AMR}, this effect is currently not measured. Such jitter could be reduced with more stable \glspl{AMR}, rigid camera and laser diode mounts, mechanical stabilizers, and by smoothing the trajectory. Replacing the crosshair center with a position reference on the \gls{AMR} which eliminates the jitter issue, or with a stabilized point in image coordinates, makes the laser diode obsolete. Using no laser diode also increases robustness to illumination changes, which affect mainly the crosshair detection and less the square detection, and prevents missed chessboard corner detections caused by overlaps with the crosshair. Overall, we believe the next version of GG should be without the laser diode, which was kept in the prototype for reduced jitter compared with using a calibrated, virtual reference point in the image.

\textbf{3D positioning:}
Future work could study \gls{PnP}-based localization, as often used with fiducial markers. The known camera intrinsics and correspondences between the detected 2D image chessboard corner points and their known 3D world positions enable 3D camera position and orientation estimation. \gls{PnP} requires no laser diode and might also reduce the error from projecting the estimated position by \gls{GG} to the position of another localization sensor on the \gls{AMR}.

\textbf{Global localization:}
By placing \gls{QR} codes on the floor, \gls{GG} achieves absolute, global localization. \gls{QR} codes were, however, just a convenient solution for this proof-of-concept. Simpler visual encodings with faster detection would improve the system, especially for online operation. The encodings serve to identify the chessboard square corners, e.g. with the known QR corners as in our current version, or by alternatives such as identifiable grid lines.

\textbf{Real-time processing:}
Instead of running the full detection chain for each image as described in Sec.~\ref{sec:ground_gazer}, one could track the global references and chessboard corners after initial detection. This requires processing only smaller image patches, thus speeding up the detection chain. Without the laser diode, planar localization mainly becomes the problem of tracking equally spaced chessboard corners, which should enable very fast position estimates. The online positioning accuracy of a runtime optimized GG might therefore be more affected by the system delay introduced by the collection of the image and its transmission to the processor rather than the processing time itself, at least in clean environments that allow for effective corner tracking. To counteract such delays, an \gls{IMU} could be integrated for higher frequency position updates.

\textbf{Robustness:}
The redundant, periodic chessboard pattern allows \gls{GG} to operate even when large areas of the chessboard are occluded, e.g. by other \glspl{AMR}, people, dirt etc. Robustness could be increased by adding more cameras and odometry, which would additionally improve accuracy, but also increase system complexity and cost.

\section{Conclusion}
We have introduced a camera-based indoor localization system called \glsreset{GG}\gls{GG} for 2D \gls{AMR} position and heading estimation. Opposed to currently existing methods with similar mm localization and sub-degree heading accuracy, our system has much lower hardware requirements. Its limitation is the need for a chessboard or a similar grid pattern and global position and orientation references on the floor, which may limit its usability for applications where this is impractical. We built a prototype of \gls{GG} using a chessboard floor with known corner coordinates as passive position references, QR codes as global position and orientation references, a ground gazing fisheye camera placed at low height on the \gls{AMR}, and a laser diode that projects a crosshair from the \gls{AMR} onto the chessboard floor. The crosshair and the floor are continuously observed by the camera for position and heading estimation. Our prototype achieved low mm localization and sub-degree heading accuracy. We have further discussed the advantages this localization approach brings and how it may be improved and extended in various directions in future work.

\bibliographystyle{ieeetr}
\bibliography{ref}
\end{document}